\documentclass[apj]{emulateapj}
\usepackage{apjfonts}
\usepackage{times}

\usepackage{amsmath}
\usepackage{graphicx}
\usepackage{natbib}
\usepackage{color}

\slugcomment{Accepted for publication in ApJ }
\shorttitle{}
\shortauthors{Shin et al.}

\newcommand{\mbh}{$M_{\rm BH}$}     
\newcommand{\zBLR}{$Z_{\rm BLR}$}     
\newcommand{\ergs}{erg~s$^{\rm -1}$}
\newcommand{\msun}{$M_{\odot}$}

\begin{document}
%\title{The mass-metallicity relation of low-redshift QSOs}
\title{The chemical properties of low-redshift QSOs}
\author{Jaejin Shin$^{1}$}
\author{Jong-Hak Woo$^{1}$\altaffilmark{,5}}
\author{Tohru nagao$^{2, 3}$}
\author{Sang Chul Kim$^{4}$}

\affil{
$^1$Astronomy Program, Department of Physics and Astronomy, 
Seoul National University, Seoul, 151-742, Republic of Korea\\
$^2$The Hakubi Center for Advanced Research, Kyoto University, 
Yoshida-Ushinomiya-cho, Sakyo-ku, Kyoto 606-8302, Japan\\
$^3$Department of Astronomy, Kyoto University,
Kitashirakawa-Oiwake-cho, Sakyo-ku, Kyoto 606-8502, Japan\\
$^4$Korea Astronomy and Space Science Institute, Daejeon 305-348, Republic of Korea
}
\altaffiltext{5}{Author to whom any correspondence should be addressed}

\begin{abstract}
We investigate the chemical properties of low-$z$ QSOs, using archival UV spectra obtained 
with the HST and IUE for a sample of 70 Palomar-Green QSOs at $z < 0.5$. 
By utilizing the flux ratio of UV emission lines (i.e., N~{\sc v} /C~{\sc iv}, 
(Si~{\sc iv}+O~{\sc iv}])/C~{\sc iv}, and N~{\sc v}/He~{\sc ii}) 
as metallicity indicators, we compare broad-line region (BLR) gas metallicity
with AGN properties, i.e., black hole mass, luminosity, and Eddington ratio.
We find that BLR metallicity correlates with Eddington ratio while the dependency 
on black hole mass is much weaker. Although these trends of low-$z$ AGNs appear to be 
different from those of high-$z$ QSOs, the difference between low-$z$ and high-$z$ samples is 
partly caused by the limited dynamical range of the samples.
We find that metal enrichment at the center of galaxies is closely connected to the accretion
activity of black holes and that the scatter of metallicity correlations with
black hole mass increases over cosmic time.

\end{abstract}

%%Section 1
\section{INTRODUCTION} \label{section:intro}

Measuring chemical properties of galaxies and their redshift evolution is a crucial 
step in understanding galaxy evolution since the metallicity of galaxies is closely 
related to the history of star-formation, gas inflow, and outflow.
A number of observational studies have been devoted to measure the metallicity of galaxies,
revealing metallicity correlations with various galaxy properties. 
In the local universe, it has been shown that metallicity scales with galaxy luminosity
and mass (the luminosity-metallicity relation and mass-metallicity relation, respectively)
based on the metallicity measured from gas emission lines \citep[][and references therein]
{tremonti2004} or stellar absorption lines \citep[e.g.,][]{Gallazzi2005, Panter2008}. 
These scaling relations indicate that metal enrichment is closely connected
to the galaxy mass assembly. 	

Recently the redshift evolution of these scaling relations in
star-forming galaxies has been extensively investigated.
At $z<3$, many studies have suggested apparent metallicity
evolution as a function of redshift. The evolution is significant especially
for low stellar-mass galaxies \citep[e.q.,][]{Savaglio2005,Erb2006,Maiolino2008,Mannucci2009,Yabe2012},
 though it could be due to observational selection effects \citep[see][]{Mannucci2010}. 
At $z>3$, exploring the metallicity of galaxies is extremely challenging, because 
typical galaxies at such 
high redshifts are very faint and the classical metallicity indicators in the 
rest-frame optical spectra shift out of the atmospheric windows \citep[but see 
also, e.g.,][]{Laskar2011, Nagao2012}.

To extend the metallicity measurement toward higher redshifts, one possible 
approach is to focus on active galactic nuclei (AGNs). Thanks to their high 
luminosity ($L_{\rm AGN}$) and a various metallic emission lines in their 
rest-frame ultraviolet spectra, it is possible to infer the metallicity of 
broad-line regions (BLRs) even for QSOs at $z \sim 6-7$ with ground-based telescopes 
(\citealt{Kurk2007, Juarez2009}; see also \citealt{Mortlock2011}). 
Previously a positive relation between the metallicity of BLRs ($Z_{\rm BLR}$) 
and the redshift of AGNs has been reported \citep[e.g.,][]{Hamann1992,hamann1993}; 
however it turned out that the apparent relation was caused by a selection bias and 
that the correlation between AGN luminosity and BLR metallicity 
(the $L_{\rm AGN} - Z_{\rm BLR}$ relation) was fundamental
(see, e.g., \citealt{Nagao2006b}).
Note that the luminosity-metallicity relation of AGNs has been also reported 
based on the emission lines in the narrow-line region (NLR), which is much more
extended than BLR and traces the chemical properties in the spatial scale of AGN host 
galaxies \citep{Nagao2006a, Matsuoka2009}. Interestingly, the 
luminosity-metallicity relation of high-luminosity QSOs shows no strong redshift 
evolution in the redshift range of $2 < z < 6$ \citep{Nagao2006a, Nagao2006b, Juarez2009, 
matsuoka2011b}, implying that the chemical evolution at the center of host galaxies
is mostly completed at a very high redshift \citep[see also][]{matsuoka2011a}.

Although the observed $L_{\rm AGN} - Z_{\rm BLR}$ relation in AGNs and its 
redshift dependence are crucial to constrain evolutionary scenarios of the 
supermassive black hole (BH), the host galaxy, and the interplay between 
these two (i.e., the galaxy-BH coevolution), there are two main drawbacks 
that should be resolved. First, while the $L_{\rm AGN} - Z_{\rm BLR}$ relation 
is well established, its origin is still controversial. 
For example, \cite{Warner2004} 
reported that the metallicity of BLR showed a correlation with the mass of BH 
(\mbh), but no correlation with the Eddington ratio ($L/L_{\rm Edd}$). 
Their result was recently confirmed by a larger sample of QSOs    
\citep{matsuoka2011b}. On the other hand, \cite{Shemmer2004} claimed that 
the observed $L_{\rm AGN} - Z_{\rm BLR}$ relation was caused by the dependence 
of  $Z_{\rm BLR}$ on the Eddington ratio, not on \mbh\ \citep[see also][]
{Dietrich2009}. At lower redshifts, it has been reported that narrow-line Seyfert 
1 galaxies \citep[NLS1s, whose Eddington ratios are believed to be high; 
see, e.g., ][]{Boroson2002,Grupe2004} 
show higher metallicity than typical broad-line AGNs \citep[e.g.,][]
{Wills1999, Nagao2002, Shemmer2002}. The higher metallicity of NLS1s in these
studies is qualitatively consistent with the result reported by \cite{Shemmer2004}. 

The other drawback comes from observational limitations. 
The AGN metallicity based on both BLRs and NLRs has been examined predominantly 
for AGNs at $z>2$, since the AGN metallicity studies generally utilizes emission lines 
in the rest-frame ultraviolet spectra. 
Consequently, the observational studies with ground-based telescopes 
are limited to AGNs at $z>2$. 
This prevents us from studying the difference in the chemical 
properties of BLRs between high-$z$ and low-$z$ QSOs. For instance, 
\cite{Nagao2006b} studied the $L_{\rm AGN} - Z_{\rm BLR}$ relation of QSOs at 
$2.0 \leq z \leq 4.5$. The possible physical origin of this $L_{\rm AGN} - 
Z_{\rm BLR}$ relation has been examined by \cite{Shemmer2004} for $2.0 < z < 
3.5$, and by \cite{matsuoka2011b} for $2.3 < z < 3.0$. Although \cite{Warner2004} 
investigate this issue for AGNs in a wide redshift range of $0 < z < 5$, they did not
examine the redshift dependence of BLR chemical properties. 
Although there are a few attempts to 
infer the NLR metallicity based on the rest-frame optical spectra \cite[e.q.,][]
{Storchi-Bergmann1998, Nagao2002, Groves2006}, those methods are in turn difficult 
to apply for high-$z$ AGNs since infrared spectroscopy is required to obtain rest-frame
optical emission lines, thus inconvenient for the comparative study 
between high and low redshifts. 
Using space observations with the $IUE$ and the $HST$, 
\cite{Shemmer2002} investigated UV spectra of low-$z$ AGNs and showed that a significant 
$L_{\rm AGN} - Z_{\rm BLR}$ relation was present also in low-$z$ AGNs. 
However, they did not examine the physical origin of the $L_{\rm AGN} - 
Z_{\rm BLR}$ relation and thus it is not clear whether BH mass or the accretion
rate derives the observed $L_{\rm AGN} - Z_{\rm BLR}$ relation, and whether
the low-$z$ and high-$z$ AGNs show the same correlations.

Motivated by these considerations, in this paper we investigate the correlation
between BLR metallicity with various AGN properties, including mass, luminosity, and
Eddington ratio, for a sample of 70 low-$z$ Palomar-Green (PG) QSO at $z < 0.5$, 
by utilizing the archival UV spectra. 
We describe the sample selection and the data in \S 2, the data analysis and the 
fitting procedure in \S 3. The main results are presented in \S 4, followed by 
discussion in \S 5 and summary and conclusions in \S 6. 
We adopt a cosmology of $H_{\rm 0}= 70$ km  s$^{-1}$ Mpc$^{-1}$,  
$\Omega_{\Lambda}=0.7$ and $\Omega_{\rm m}=0.3$.

%section 2
\section{Sample Selection and the Data} \label{section:sample}

%section 2.1
\subsection{Sample selection} \label{section:meas:MgII}
To investigate metallicity of high-$z$ QSOs, both permitted and weak semi-forbidden lines
have been used. The weak semi-forbidden emission lines in the rest-frame UV spectra, i.e., 
N~{\sc iv}]$\lambda$1486, O~{\sc iii}]$\lambda$1663, and N~{\sc iii}]$\lambda$1749 
are good metallicity indicators
\cite[e.g.,][]{Shields1976,baldwin1978,Osmer1980,Uomoto1984,Warner2002},
since the flux ratio among these lines do not show strong dependences on physical properties 
of gas clouds (such as the density and ionization parameter). However, 
the actual application to observational data is generally difficult 
since those semi-forbidden emission lines are too faint to be measured accurately. 
Thus, stronger emission lines are preferred in the studies of $Z_{\rm BLR}$.
For example, the flux ratios of N~{\sc v}$\lambda$1240 to C~{\sc iv}$\lambda$1549 and
N~{\sc v}$\lambda$1240 to He~{\sc ii}$\lambda$1640 have been utilized to infer 
$Z_{\rm BLR}$ by comparing them with the prediction of photoionization models \cite[e.q.,][]
{Hamann1992,hamann1993,Ferland1996,Korista1998,Dietrich1999,Dietrich2000,
Hamann2002,Dietrich2003}. \cite{Nagao2006b} showed that the flux ratios of 
(Si~{\sc iv}$\lambda$1397+O~{\sc iv}]$\lambda$1402)/C~{\sc iv}$\lambda$1549
and Al~{\sc iii}$\lambda$1857/C~{\sc iv}$\lambda$1549 are also useful to infer
$Z_{\rm BLR}$ through photoionization model runs, that are used for inferring
$Z_{\rm BLR}$ in high-$z$ QSOs \citep[e.g.,][]{Juarez2009, matsuoka2011b}. 

Since most previous studies used the rest-frame UV spectra to infer BLR metallicity
of high-redshift Type-1 QSOs, we therefore focus on high-luminosity QSOs at 
low-redshfit, for which UV spectra are available, in order to investigate the chemical
properties of low-$z$ QSOs compared to high-$z$ QSOs. 
The Palomar-Green (PG) QSOs \citep{Schmidt1983} are well-studied low-$z$ luminous Type-1 AGNs,
and the UV spectra of many PG QSOs have been previously obtained with space facilities,
%i.e., the $HST$ and the $IUE$, 
thus suitable for our BLR metallicity study.

We selected all PG QSOs at $z < 0.5$ (89 objects), for which reliable black hole masses are available 
from either reverberation mapping results \citep{Peterson2004,Denney2010}  or single-epoch 
estimates \citep{VP06}. 
To investigate BLR metallicity, we will use the flux ratios of 
N~{\sc v}$\lambda$1240/C~{\sc iv}$\lambda$1549,
N~{\sc v}$\lambda$1240/He~{\sc ii}$\lambda$1640, and
(Si~{\sc iv}$\lambda$1397+O~{\sc iv}]$\lambda$1402)/C~{\sc iv}$\lambda$1549,
since these emission lines have relatively large equivalent widths.
Thus, we searched for available UV spectra previously obtained from space facilities,
using the Mikulski Archive for Space Telescopes (MAST).
Among 89 PG QSOs at $z < 0.5$, the archival UV spectra covering the required
emission lines were available for 86 objects. Among them, 
we excluded 7 broad absorption 
line (BAL) QSOs, since the strong absorption features prevent us from 
measuring the emission-line fluxes accurately. We also excluded 9 additional 
objects, for which the spectral quality is too low to identify the aforementioned
emission lines. Thus, we finalized a sample of 70 PG QSOs for this work as 
listed in Table 1.

\begin{deluxetable*}{lcllclcr} 
\tablewidth{0pt}
\tablecolumns{8}
\tabletypesize{\scriptsize}
\tablecaption{Log of archival UV data \& AGN properties}

\tablehead{
\colhead{Object} &
\colhead{Redshift} &
\colhead{Observation Date} &
\colhead{Telescope/Instrument} &
\colhead{log[$M_{\rm BH}/M_{{\odot}}$]} &
\colhead{Ref.}&
\colhead{log[$L_{\rm bol}/\rm erg\rm s^{- 1}$]}&
\colhead{S/N}
\\
\colhead{(1)} &
\colhead{(2)} &
\colhead{(3)} &
\colhead{(4)} &
\colhead{(5)} &
\colhead{(6)} &
\colhead{(7)} &
\colhead{(8)} 
}

\startdata
PG0003+158	& $0.450$	  &1993 Nov 05,07 		&HST/FOS	& $	9.25	\pm	0.03	$&   2  	&$	46.64	\pm	0.18	$&$	5.47	$	 \\
PG0003+199	& $0.026$  & 2010 Feb 08			&HST/COS	& $	7.13	\pm	0.11	$&   1	&$	44.43	\pm	0.04	$&$	28.09	$	 \\
PG0007+106	& $0.089$  &1981 Jun 08 			&IUE/SWP	& $	8.71	\pm	0.09	$&   2  	&$	44.96	\pm	0.10	$&$	7.30	$	 \\
PG0026+129	& $0.142$  &1994 Nov 27			&HST/FOS	& $	8.57	\pm	0.11	$&   1	&$	45.57	\pm	0.03	$&$	34.59	$	 \\
PG0049+171	& $0.064$  &1985 Jul 31			&IUE/SWP	& $	8.33	\pm	0.09	$&   2  	&$	44.36	\pm	0.23	$&$	3.22	$	 \\
PG0050+124	& $0.061$  &1979 Dec 22,23			&IUE/SWP	& $	7.42	\pm	0.10	$&   2  	&$	44.69	\pm	0.06	$&$	12.12	$	 \\
PG0052+251	& $0.155$  &1992 Jun 29			&IUE/SWP	& $	8.55	\pm	0.09	$&   1	&$	45.78	\pm	0.05	$&$	14.40	$	 \\
PG0157+001	& $0.164$  &1985 Aug 09			&IUE/SWP	& $	8.15	\pm	0.09	$&   2  	&$	45.70	\pm	0.07	$&$	10.30	$	 \\
PG0804+761	& $0.100$  &2010 Jun 12			&HST/COS	& $	8.82	\pm	0.05	$&   1	&$	45.99	\pm	0.02	$&$	62.65	$	 \\
PG0838+770	& $0.131$  & 2009 Sep 24			&HST/COS	& $	8.13	\pm	0.09	$&   2  	&$	45.24	\pm	0.06	$&$	19.15	$ \\
PG0844+349	& $0.064$  &1987 Nov 30;Dec 01			&IUE/SWP	& $	7.95	\pm	0.18	$&   1	&$	45.00	\pm	0.06	$&$	12.63	$	 \\
PG0921+525	& $0.035$  &1988 Feb 28,29			&IUE/SWP	& $	7.38	\pm	0.11	$&   1	&$	44.30	\pm	0.05	$&$	14.94	$	 \\
PG0923+129	& $0.029$  &1985 May 01			&IUE/SWP	& $	8.58	\pm	0.10	$&   2  	&$	44.14	\pm	0.08	$&$	9.91	$	 \\
PG0947+396	& $0.206$  &1996 May 06			&HST/FOS	& $	8.66	\pm	0.09	$&   2  	&$	45.84	\pm	0.14	$&$	6.78	$	 \\
PG1011--040	& $0.058$  &2010 Mar 26			&HST/COS	& $	7.30	\pm	0.09	$&   2  	&$	44.83	\pm	0.03	$&$	35.03	$	 \\
PG1012+008	& $0.185$  &1990 Apr 10,			&IUE/SWP	& $	8.23	\pm	0.09	$&   2  	&$	45.41	\pm	0.10	$&$	7.46	$	 \\
PG1022+519	& $0.045$  &1983 May 31;Jun 01			&IUE/SWP	& $	6.32	\pm	0.19	$&   1	&$	44.48	\pm	0.09	$&$	8.53	$	 \\
PG1048+342	& $0.167$  &1993 Nov 13			&IUE/SWP	& $	8.35	\pm	0.09	$&   2  	&$	44.74	\pm	0.68	$&$	1.12	$	 \\
PG1049-005	& $0.357$  &1992 Apr 01,			&HST/FOS	& $	9.16	\pm	0.09	$&   2  	&$	46.17	\pm	0.22	$&$	4.36	$	 \\
PG1103--006	& $0.425$  &1992 Dec 29			&HST/FOS	& $	9.30	\pm	0.10	$&   2  	&$	46.11	\pm	0.11	$&$	8.68	$	 \\
PG1115+407	& $0.154$  &1996 May 19			&HST/FOS	& $	7.65	\pm	0.09	$&   2  	&$	45.62	\pm	0.08	$&$	12.30	$	 \\
PG1116+215	& $0.177$  &1993 Feb 19,20			&HST/FOS	& $	8.51	\pm	0.09	$&   2  	&$	46.30	\pm	0.15	$&$	6.56	$	 \\
PG1119+120	& $0.049$  &1982 Nov 21,26			&IUE/SWP	& $	7.45	\pm	0.09	$&   2  	&$	44.62	\pm	0.06	$&$	12.89	$	 \\
PG1121+422	& $0.234$  &1995 Jan 08			&IUE/SWP	& $	8.01	\pm	0.09	$&   2  	&$	45.94	\pm	0.11	$&$	7.06	$	 \\
PG1149--110	& $0.049$  &1992 Dec 29			&IUE/SWP	& $	7.90	\pm	0.10	$&   2  	&$	44.25	\pm	0.10	$&$	7.58	$	 \\
PG1151+117	& $0.176$  &1987 Jan 29,30			&IUE/SWP	& $	8.53	\pm	0.09	$&   2  	&$	45.65	\pm	0.11	$&$	6.68	$	 \\
PG1202+281	& $0.165$  &1996 Jul 21			&HST/FOS	& $	8.59	\pm	0.09	$&   2  	&$	44.95	\pm	0.14	$&$	7.07	$	 \\
PG1211+143	& $0.085$  &2002 Feb 04,07			&HST/STIS	& $	7.94	\pm	0.09	$&   2  	&$	45.63	\pm	0.04	$&$	19.41	$	 \\
PG1216+069	& $0.334$  &1993 Mar 15			&HST/FOS	& $	9.18	\pm	0.09	$&   2  	&$	46.52	\pm	0.14	$&$	7.03	$	 \\
PG1226+023	& $0.158$  &1991 Jul 9			&HST/FOS	& $	8.93	\pm	0.09	$&   1	&$	46.59	\pm	0.09	$&$	10.60	$	 \\
PG1229+204	& $0.063$  &1982 May 01,02			&IUE/SWP	& $	7.84	\pm	0.21	$&   1	&$	45.11	\pm	0.04	$&$	21.02	$	 \\
PG1244+026	& $0.048$  &1983 Feb 08			&IUE/SWP	& $	6.50	\pm	0.09	$&   2  	&$	44.30	\pm	0.10	$&$	7.72	$	 \\
PG1259+593	& $0.472$  &1991 Dec 27			&HST/FOS	& $	8.90	\pm	0.10	$&   2  	&$	46.76	\pm	0.23	$&$	4.15	$	 \\
PG1302-102	& $0.286$  &1986 Jul 25,26			&IUE/SWP, LWP	& $	8.86	\pm	0.10	$&   2  	&$	46.45	\pm	0.01	$&$	57.25	$	 \\
PG1307+085	& $0.155$  &1980 May 04			&IUE/SWP	& $	8.62	\pm	0.12	$&   1	&$	45.80	\pm	0.07	$&$	10.33	$	 \\
PG1310--108	& $0.035$  &1995 Feb 11			&IUE/SWP	& $	7.86	\pm	0.09	$&   2  	&$	44.13	\pm	0.10	$&$	7.66	$	 \\
PG1322+659	& $0.168$  &1997 Jan 19			&HST/FOS	& $	8.26	\pm	0.11	$&   2  	&$	45.52	\pm	0.04	$&$	23.02	$	 \\
PG1341+258	& $0.087$  &1995 Mar 22			&IUE/SWP	& $	8.02	\pm	0.10	$&   2  	&$	44.71	\pm	0.13	$&$	5.70	$	 \\
PG1351+695	& $0.030$  &2011 Jun 27			&HST/COS	& $	7.52	\pm	0.12	$&   1	&$	43.63	\pm	0.10	$&$	8.61	$	 \\
PG1352+183	& $0.158$  &1996 May26			&HST/FOS	& $	8.40	\pm	0.09	$&   2  	&$	45.60	\pm	0.11	$&$	8.52	$	 \\
PG1402+261	& $0.164$  &1996 Aug 25			&HST/FOS	& $	7.92	\pm	0.09	$&   2  	&$	45.95	\pm	0.08	$&$	12.85	$	 \\
PG1404+226	& $0.098$  &1996 Feb 23			&HST/FOS	& $	6.87	\pm	0.09	$&   2  	&$	44.86	\pm	0.15	$&$	6.62	$	 \\
PG1415+451	& $0.114$  &1997 Jan 02 			&HST/FOS	& $	7.99	\pm	0.09	$&   2  	&$	45.29	\pm	0.08	$&$	12.11	$	 \\
PG1416--129	& $0.129$  &1988 Mar 03			&IUE/SWP	& $	9.02	\pm	0.09	$&   2  	&$	44.93	\pm	0.17	$&$	4.51	$	 \\
PG1425+267	& $0.366$  &1996 Jun29			&HST/FOS	& $	9.71	\pm	0.11	$&   2  	&$	46.15	\pm	0.06	$&$	17.63	$	 \\
PG1426+015	& $0.086$  &2004 Jul 27, 28, 29			&HST/STIS	& $	9.09	\pm	0.13	$&   1	&$	45.63	\pm	0.07	$&$	10.18	$	 \\
PG1427+480	& $0.221$  &1997 Jan 07			&HST/FOS	& $	8.07	\pm	0.09	$&   2  	&$	45.75	\pm	0.10	$&$	9.61	$	 \\
PG1434+590	& $0.031$  &2009 Aug 04			&HST/COS	& $	7.77	\pm	0.12	$&   1*	&$	44.93	\pm	0.03	$&$	37.65	$	 \\
PG1435--067	& $0.129$  &1995 Jun 12			&IUE/SWP	& $	8.34	\pm	0.09	$&   2  	&$	45.56	\pm	0.09	$&$	8.74	$	 \\
PG1440+356	& $0.077$  &1996 Dec 05			&HST/FOS	& $	7.45	\pm	0.09	$&   2  	&$	45.59	\pm	0.06	$&$	17.75	$	 \\
PG1444+407	& $0.267$  &1996 May 23			&HST/FOS	& $	8.27	\pm	0.09	$&   2  	&$	46.24	\pm	0.10	$&$	9.54	$	 \\
PG1448+273	& $0.065$  &2011 Jun 18	 	 	&HST/COS	& $	6.95	\pm	0.09	$&   2  	&$	44.36	\pm	0.03	$&$	13.07	$	\\
PG1501+106	& $0.036$  &1989 Jun 30; Jul 02 			&IUE/SWP	& $	8.50	\pm	0.09	$&   2  	&$	44.51	\pm	0.03	$&$	29.62	$	 \\
PG1512+370	& $0.371$  &1992 Jan 26			&HST/FOS	& $	9.35	\pm	0.09	$&   2  	&$	46.36	\pm	0.17	$&$	5.66	$	 \\
PG1519+226	& $0.137$  &1995 Jun 11			&IUE/SWP	& $	7.92	\pm	0.09	$&   2  	&$	45.16	\pm	0.18	$&$	4.27	$	 \\
PG1534+580	& $0.030$  &2009 Oct 28			&HST/COS	& $	7.37	\pm	0.07	$&   1*	&$	44.16	\pm	0.05	$&$	22.45	$	 \\
PG1543+489	& $0.400$  &1995 Mar 14			&HST/FOS	& $	7.98	\pm	0.09	$&   2  	&$	46.26	\pm	0.04	$&$	28.04	$	 \\
PG1545+210	& $0.266$  &1992 Apr 08,10			&HST/FOS	& $	9.29	\pm	0.09	$&   2  	&$	45.98	\pm	0.10	$&$	10.04	$	 \\
PG1552+085	& $0.119$  &1986 Apr 28			&IUE/SWP	& $	7.52	\pm	0.09	$&   2  	&$	44.81	\pm	0.24	$&$	3.14	$	 \\
PG1612+261	& $0.131$  &1980 Sep 10			&IUE/SWP	& $	8.04	\pm	0.09	$&   2  	&$	45.07	\pm	0.15	$&$	5.13	$	 \\
PG1613+658	& $0.129$  &2010 Apr 08, 09, 10			&HST/COS	& $	8.43	\pm	0.20	$&   1	&$	45.94	\pm	0.02	$&$	53.54	$	 \\
PG1617+175	& $0.112$  &1993 May 13			&IUE/SWP	& $	8.75	\pm	0.10	$&   1	&$	45.24	\pm	0.09	$&$	8.66	$	 \\
PG1626+554	& $0.133$  &1997 Nov 19			&HST/FOS	& $	8.48	\pm	0.09	$&   2  	&$	45.72	\pm	0.07	$&$	14.30	$	 \\
PG2112+059	& $0.466$  &1992 Sep 19			&HST/FOS	& $	8.98	\pm	0.10	$&   2  	&$	46.25	\pm	0.18	$&$	5.53	$	 \\
PG2130+099	& $0.063$  &2010 Oct 28,			&HST/COS	& $	8.64	\pm	0.05	$&   1	&$	44.92	\pm	0.03	$&$	34.83	$ \\
PG2214+139	& $0.067$  &1984 Jun 03			&IUE/SWP	& $	8.53	\pm	0.10	$&   2  	&$	44.39	\pm	0.84	$&$	0.89	$	 \\
PG2233+134	& $0.325$  &2003 May 13			&HST/STIS	& $	8.02	\pm	0.09	$&   2  	&$	46.16	\pm	0.04	$&$	21.26	$	 \\
PG2251+113	& $0.323$  &2001 May 01			&HST/STIS	& $	8.97	\pm	0.09	$&   2  	&$	45.83	\pm	0.05	$&$	14.15	$	 \\
PG2304+042	& $0.042$  &1989 Dec 29			&IUE/SWP	& $	8.54	\pm	0.10	$&   2  	&$	43.72	\pm	0.25	$&$	3.03	$	 \\
PG2308+098	& $0.432$  &1992 Oct 12			&HST/FOS	& $	9.57	\pm	0.11	$&   2  	&$	46.33	\pm	0.18	$&$	5.38	$	 \\
\enddata
\label{table:prop}

\tablecomments{
    Col. (1): Target ID. 
    Col. (2): Redshift. 
    Col. (3): Observed date. 
    Col. (4): Telescope and Instrument. 
    Col. (5): Black hole mass from \cite{Peterson2004,VP06,Denney2010} with a new virial factor \citep{woo2010}. 
    Col. (6): References for redshift and black hole mass. 
        1 - Reverberation-mapped AGNs \citep{Peterson2004},
        1* -  Reverberation-mapped AGNs \citep{Denney2010},
        2 - AGNs with single-epoch black hole mass \citep{VP06}.
    Col. (7): AGN bolometric luminosity calculated from the monochromatic luminosity
at 1350\AA\ by multiplying a bolometric correction factor, 3.81. 
    Col. (8): Signal-to-noise ratio per resolution element at 1350\AA\ in the rest-frame. 
  }
\end{deluxetable*}

%section 2.2
\subsection{Data} \label{section:Data}

We obtained the UV spectra taken
with {\it International Ultraviolet Explorer} ($IUE$) or {\it Hubble Space Telescope}
($HST$) through the MAST database. 
We collected all available spectra of our targets and used the best quality spectrum 
with an order of {\it Cosmic Origins Spectrograph} ($COS$), {\it Space Telescope Imaging Spectrograph} ($STIS$), {\it Faint Object Spectrograph} ($FOS$), and $IUE$ when
multiple instruments have been used. In summary, we utilized 10 COS spectra, 
4 STIS spectra, 26 for FOS and 30 $IUE$ spectra.  

Specifically, we used the SWP ($1200-2000$\AA) data of $IUE$, G130H ($1150-1600$\AA) and G160H 
($1600-2300$\AA) data of $HST$/FOS, and G140M ($1150-1740$\AA) data of 
$HST$/STIS, for lower-$z$ QSOs. For relatively higher-$z$ QSOs, we used 
the LWP ($1800-3200$\AA) data of $IUE$ and G270H ($2300-3200$\AA) data of 
$HST$/FOS data. Finally,  for the $HST$/COS data, we used the combined data
in two spectral ranges, i.e., G130M ($1150-1450$\AA) and G160M ($1405-1775$\AA),
in order to cover the N~{\sc v} and C~{\sc iv} lines at the same time.

%If there are some available spectral data in one observing run, we 
%used all of the available data by combining and connecting them. 
We combined the spectra of each exposures by calculating the error-weighted 
mean. For the STIS data, we combined the spectra using the exposure-time 
as a weight, because we could not eliminate artificial spark features 
effectively when we adopted the error-weighted mean.
However, the spectra are qualitatively consistent.
In the case of COS spectra, we used IDL routines developed by the COS GTO team \citep{danforth2010}.
We smoothed the spectra in the wavelength 
direction by adopting 7 pixel smoothing for the COS data and 2 pixel smoothing for the 
STIS data. 

If a target has been observed at multiple-epochs, 
we chose only one epoch with the best data quality to avoid any time-variation effects. 
Table 1 lists the observation data and the instrument for each target.
In this table we also list the signal-to-noise ratio
per resolution element calculated at the rest-frame 1350\AA\ continuum.

%section 2.3
\subsection{AGN properties} \label{section:AGN properties}
To compare with BLR metallicity, we measure and collect other AGN properties, i.e.,
black hole mass, bolometric luminosity, and Eddington ratio.
We collected black hole mass of the sample QSOs, which has been previously determined 
by the reverberation mapping studies for 18 objects \citep{Peterson2004,Denney2010} or by 
the single-epoch method for 52 objects \citep{VP06}.
A black hole mass measurement based on reverberation-mapping results
is availble for PG1211+143 \citep{Peterson2004}, however it has large uncertainty
due to the low data quality, and it has been excluded in other reverberation 
sample studies. Thus, we will use single-epoch mass for PG1211+143.

We re-calculated black hole mass of the sample by adopting the updated virial factor of 5.2 \citep{woo2010}, 
which is slightly smaller than the previous virial factor (5.5; \citealt{Onken2004}, see also \citealt{Park2012})

As the uncertainty of black holes masses, we adopted the values given by \cite{Peterson2004,VP06,Denney2010}. 

For AGN bolometric luminosity ($L_{\rm AGN}$), we used the obtained UV spectra
to measure monochromatic luminosity at 1350\AA, which is presumably not heavily 
contaminated by the host galaxy stellar light.  
To measure the flux at 1350\AA, we fitted the AGN continuum between 1210\AA\ and 1700\AA\ with a
power-law function. The measured 1350\AA\ monochromatic continuum luminosity is then
used for calculating AGN bolometric luminosity by multiplying a bolometric correction factor, 3.81 \citep{shen2008}. 
Note that this bolometric correction factor
is the same as adopted by \cite{matsuoka2011b} for high-$z$ QSOs. %(see Section 5). 
The measurement uncertainty of AGN luminosities was calculated based on the signal-to-noise ratio of the spectra. 

In Figure 1, we present the distribution of the sample properties; the redshift, black hole mass,
bolometric luminosity, and Eddington ratio.  
The black hole mass ranges over 3 orders of magnitude (from 6.32 to 9.71) with an average
of $8.26\pm 0.71$ \msun. The bolometric luminosity also ranges over a large range from 10$^{43.6}$ to 10$^{46.8}$ \ergs while some fraction of the sample has relatively low luminosity 
and belong to Seyfert class rather than QSOs. The mean Eddington ratio of the sample is $\sim$10\% with 0.66 dex dispersion,
indicating that there is a large range of accretion activity \citep{Woo2002}.

\begin{figure}
\includegraphics[width = 0.48\textwidth]{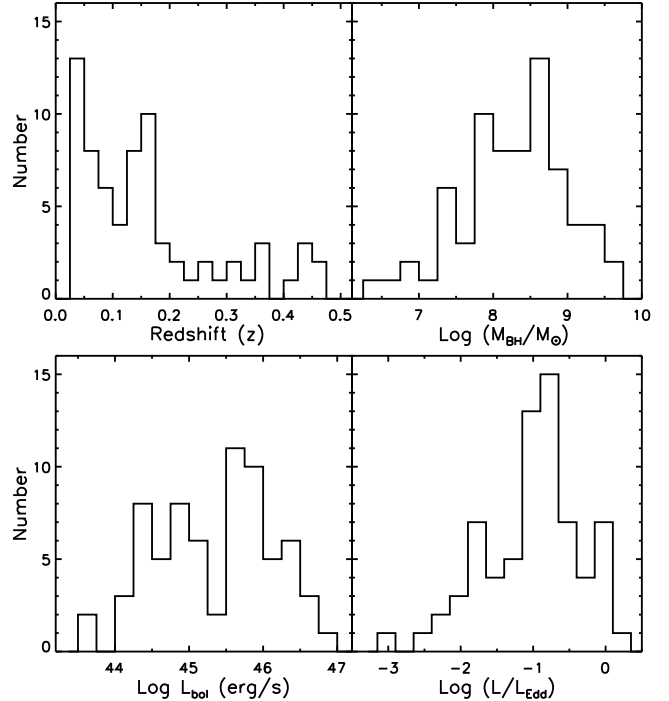}
\caption{
     The distributions of the redshift, BH mass, AGN luminosity, and the 
     Eddington ratio of the sample. 
\label{fig:allspec1}}
\end{figure}

%section 3
\section{Analysis}\label{section:analysis}

%section 3.1
\subsection{Multi-component fitting procedure} \label{section:MC fitting}

The flux measurement of BLR emission lines is an important step for investigating
$Z_{\rm BLR}$. It is known that BLR emission lines sometimes show significant
asymmetric velocity profiles \citep[e.q.,][]{corbin1997,vandenberk2001,baskin2005},
hence a single-Gaussian model does not generate a reliable fit for such cases.  
To fit asymmetric velocity profiles of QSO UV emission lines, various 
models have been adopted. Here, we examined 4 models, namely, 
double-Gaussian, Gauss-Hermitian, modified-Lorentzian, and 2 power-law functions 
to determine the best line profile to use.
In Figure 2, we present the C~{\sc iv} line of PG0003+158 as well as 4 different model
fits, which show slightly different results, particularly at the wing of the line. 
Through our visual inspection, we decided to adopt the double-Gaussian function
as an emission line profile model. Note that our results do not significantly
depend on the choice of the model since the difference in flux measurements
is $\sim$ 5\%.

\begin{figure}
	\includegraphics[width = 0.45\textwidth]{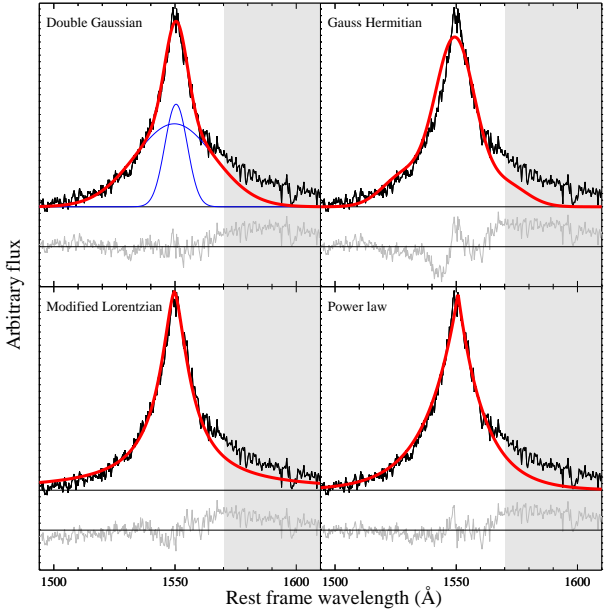}
	\centering
	\caption{Comparison of different fitting functions for the C~{\sc iv}$\lambda$1549
	emission line of PG 0003+158. Double-Gaussian (upper left), 
	Gauss-Hermitian (upper right), modified-Lorentzian (lower left) and 2 
	power-law (lower right) functions are examined. The red lines denote the fitting results,
	and the blue line in the upper left panel represents each Gaussian component.
	Residual spectrum is shown at the bottom in each panel. Masked regions are 
	indicated with gray hatches. 
	\label{fig:allspec1}}
\end{figure}

In the rest-frame UV spectra, many AGN emission lines are blended;
Ly$\alpha$ $\lambda$1216+N~{\sc v}$\lambda$1240, 
Si~{\sc iv}$\lambda$1397+O~{\sc iv}]$\lambda$1402, and 
He~{\sc ii}$\lambda$1640+O~{\sc iii}]$\lambda$1663+Al~{\sc ii}$\lambda$1671. 
Thus, it is necessary to perform a multi-component fitting analysis for 
secure flux measurements.
We simultaneously fitted all 10 emission lines, that were used as BLR metallicity indicators.
First, we divided these emission lines into two groups based on their ionization
degree, and assumed that the emission lines in each group have the same velocity 
profile \citep[see][]{Nagao2006b}. Specifically, we categorized 
N~{\sc v}$\lambda$1240, O~{\sc iv}]$\lambda$1402, N~{\sc iv}]$\lambda$1486, 
C~{\sc iv}$\lambda$1549 and He~{\sc ii}$\lambda$1640) in the high-ionization
group, while Si~{\sc ii}$\lambda$1263, Si~{\sc iv}$\lambda$1397, 
O~{\sc iii}]$\lambda$1663 and Al~{\sc ii}$\lambda$1671 in the low-ionization group,
following \cite{Nagao2006b}. 
We adopted the same velocity width and the velocity shift for each group. 
We excluded the spectral range between 1570\AA\ and 1631\AA\ from the fit, 
since an unidentified emission feature is reported in this range \citep[see, e.g.,][]{Wilkes1984,Boyle1990,Laor1994,Nagao2006b}.
In the case of Ly$\alpha$, the absorption by the intergalactic matter (IGM) 
affects the line profile significantly, particularly below 1210\AA.
Thus, we treated Ly$\alpha$ separately,
by allowing the velocity dispersion and velocity shift to be free. 
We excluded the spectral range below 1210\AA\ from the fitting procedure, because 
of the IGM absorption. 
%In the fitting procedure, we masked spectral ranges which have artificial or spark features. 
For the continuum fitting, we used a power-law and determined the slope
by using three spectral windows
(1345\AA\ -- 1355\AA, 1445\AA\ -- 1455\AA, and 1687\AA\ -- 1697\AA),
where no strong emission lines are present.

\begin{figure}
	\includegraphics[width = 0.45 \textwidth]{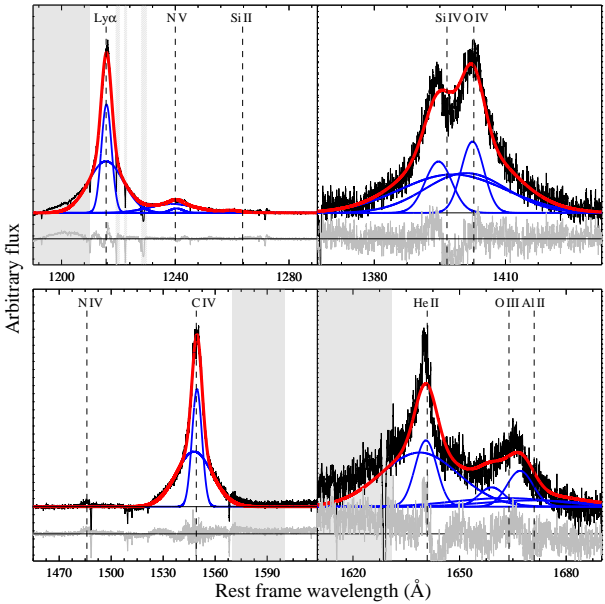}
	\centering
	\caption{{Examples of Multi-component fitting analysis. Each panel shows
	Ly$\alpha$ $\lambda$1216+N~{\sc v}$\lambda$1240 (upper left), 
 Si~{\sc iv}$\lambda$1397+O~{\sc iv}]$\lambda$1402 (upper right), 
 N~{\sc iv}]$\lambda$1486+C~{\sc iv}$\lambda$1549 (bottom left), and
 He~{\sc ii}$\lambda$1640+O~{\sc iii}]$\lambda$1663+Al~{\sc ii}$\lambda$1671 (bottom right),
 respectively.
 The color of lines are the same as in Figure 2. The dashed lines indicate the center of 
 each emission line.}
	\label{fig:allspec1}}
\end{figure}

Figure 3 shows examples of multi-component fitting for 4 different spectral
regions: from upper left to bottom right, 1) Ly$\alpha$ $\lambda$1216+N~{\sc v}$\lambda$1240, 
2) Si~{\sc iv}$\lambda$1397+O~{\sc iv}]$\lambda$1402, 3) N~{\sc iv}]$\lambda$1486+C~{\sc iv}$\lambda$1549, and
4) He~{\sc ii}$\lambda$1640+O~{\sc iii}]$\lambda$1663+Al~{\sc ii}$\lambda$1671.
The line center is denoted with dashed lines.

%section 3.2			
\subsection{Fitting result} \label{section: fitting result}

\begin{figure}
	\includegraphics[width = 0.46\textwidth]{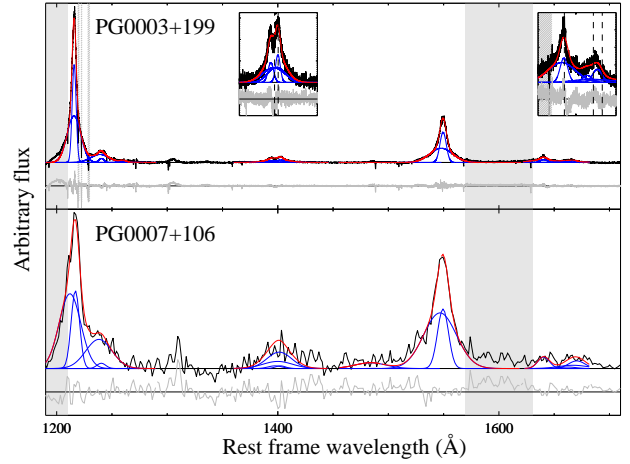}
	\caption{Fitting results of 2 representative objects: PG 0003+018 with a high
        quality COS spectrum (top) and PG 0007+106 with a low quality IUE spectrum (bottom).
        The inset panels show the fit for weak lines, namely, 
        Si~{\sc iv}+O~{\sc iv}] (left) and He~{\sc ii}+O~{\sc iii}]+Al {\sc ii} 
	(right). No inset panel means that fluxes of these weak lines were not measured 
        due to the low signal-to-noise ratio. The green dashed lines in the 
        inset panels represent the center of each line as shown in Figure 3.
        The fitting residual is shown in the lower panel for each object.
	The color of lines and masked regions are the same as in Figure 2.
	\label{fig:fig3}}
\end{figure}

Based on the multi-component fitting using double-Gaussian profiles, we 
measured the flux of 10 broad emission lines in the rest-frame UV spectra for
the sample. In Figure 4, we present the fitting results of 2 
representative targets; PG 0003+199 with high quality COS data and 
PG 0007+106 with low quality IUE data. In some cases, spectral quality
is too low to fit the weak emission lines, i.e., N~{\sc v}$\lambda$1240, 
Si~{\sc iv}$\lambda$1397, O~{\sc iv}]$\lambda$1402, C~{\sc iv}$\lambda$1549, 
and He~{\sc ii}$\lambda$1640, thus we only measure the flux of N~{\sc v} and C~{\sc iv}. 
The weak lines (i.e., Si~{\sc iv}+O~{\sc iv}] and He~{\sc ii}+O~{\sc iii}]+Al~{\sc ii}) are shown in the inset panels only when these lines were successfully fitted. 
In summary, we measured the N~{\sc v} and C~{\sc iv} fluxes for the entire sample
(70 objects) while we measured the flux of Si~{\sc iv}, O~{\sc iv}], He~{\sc ii} 
for a subsample of 34 objects. Table 2 lists the measured fluxes and the inferred uncertainties, which were estimated by averaging the signal-to-noise ratio of each pixel 
within the spectral range of each lines.
We list the sum of Si~{\sc iv}$\lambda$1397 and O~{\sc iv}]$\lambda$1402 fluxes 
instead of individual flux measurements, since the sum of two lines will be used
to compare with the flux of the C~{\sc iv} line. 

\begin{deluxetable}{lcccc} 
\tablewidth{0pt}
\tablecolumns{5}
\tabletypesize{\scriptsize}
\tablecaption{Measurements of emission Line fluxes}
\tablehead{
\colhead{Object} &
\colhead{N V} &
\colhead{Si IV+O IV]} &
\colhead{C IV}&
\colhead{He II}
\\
\colhead{} &
\colhead{} &
\colhead{(10$^{-14}$ \ergs)} &
\colhead{} &
\colhead{} 
\\
\colhead{(1)} &
\colhead{(2)} &
\colhead{(3)} &
\colhead{(4)} &
\colhead{(5)} 
}

\startdata
PG0003+158	&$	32.9	\pm	5.6	$&$	10.6	\pm	1.2	$&$	70.0	\pm	8.7	$&$	8.9	\pm	1.2	$\\
PG0003+199	&$	90.4	\pm	2.9	$&$	52.7	\pm	1.1	$&$	280.8	\pm	12.1	$&$	39.6	\pm	1.9	$\\
PG0007+106	&$	34.2	\pm	2.6	$&$	{--}	 	 	$&$	104.5	\pm	6.0	$&$	{--}	 	 	$\\
PG0026+129	&$	48.0	\pm	1.1	$&$	{--}	 	 	$&$	57.2	\pm	2.9	$&$	6.3	\pm	0.3	$\\
PG0049+171	&$	16.1	\pm	3.9	$&$	{--}	 	 	$&$	126.5	\pm	30.8	$&$	{--}	 	 	$\\
PG0050+124	&$	56.0	\pm	3.0	$&$	26.0	\pm	1.5	$&$	47.0	\pm	2.6	$&$	{--}	 	 	$\\
PG0052+251	&$	59.4	\pm	3.0	$&$	{--}	 	 	$&$	120.2	\pm	3.9	$&$	5.2	\pm	0.2	$\\
PG0157+001	&$	29.2	\pm	2.1	$&$	{--}	 	 	$&$	64.0	\pm	2.8	$&$	3.8	\pm	0.2	$\\
PG0804+761	&$	214.1	\pm	3.0	$&$	98.2	\pm	1.2	$&$	240.0	\pm	5.8	$&$	{--}	 	 	$\\
PG0838+770	&$	14.6	\pm	0.4	$&$	10.2	\pm	0.6	$&$	32.7	\pm	2.4	$&$	{--}	 	 	$\\
PG0844+349	&$	69.0	\pm	3.9	$&$	{--}	 	 	$&$	112.5	\pm	6.6	$&$	{--}	 	 	$\\
PG0921+525	&$	39.7	\pm	1.9	$&$	46.9	\pm	2.8	$&$	292.0	\pm	12.9	$&$	13.7	\pm	0.5	$\\
PG0923+129	&$	66.3	\pm	5.0	$&$	{--}	 	 	$&$	183.3	\pm	15.8	$&$	18.4	\pm	1.3	$\\
PG0947+396	&$	32.9	\pm	2.0	$&$	11.5	\pm	0.7	$&$	59.5	\pm	2.6	$&$	7.3	\pm	0.3	$\\
PG1011--040		&$	27.8	\pm	0.8	$&$	14.6	\pm	0.4	$&$	38.8	\pm	2.4	$&$	5.6	\pm	0.4	$\\
PG1012+008	&$	8.6	\pm	0.9	$&$	17.7	\pm	1.5	$&$	22.2	\pm	1.6	$&$	{--}	 	 	$\\
PG1022+519	&$	12.2	\pm	1.1	$&$	{--}	 	 	$&$	45.7	\pm	4.5	$&$	{--}	 	 	$\\
PG1048+342	&$	12.4	\pm	2.9	$&$	{--}	 	 	$&$	19.1	\pm	3.4	$&$	{--}	 	 	$\\
PG1049-005		&$	29.7	\pm	7.3	$&$	11.3	\pm	1.9	$&$	41.6	\pm	6.4	$&$	4.7	\pm	0.7	$\\
PG1103-006		&$	13.8	\pm	1.6	$&$	{--}	 	 	$&$	13.8	\pm	0.9	$&$	0.7	\pm	0.1	$\\
PG1115+407		&$	18.5	\pm	1.2	$&$	{--}	 	 	$&$	34.4	\pm	2.0	$&$	{--}	 	 	$\\
PG1116+215		&$	148.5	\pm	20.5	$&$	78.3	\pm	14.4	$&$	210.9	\pm	28.1	$&$	17.1	\pm	2.2	$\\
PG1119+120		&$	52.2	\pm	2.6	$&$	18.4	\pm	1.3	$&$	73.2	\pm	4.3	$&$	8.0	\pm	0.4	$\\
PG1121+422		&$	14.5	\pm	1.9	$&$	{--}	 	 	$&$	47.8	\pm	3.5	$&$	{--}	 	 	$\\
PG1149--110		&$	17.1	\pm	1.4	$&$	{--}	 	 	$&$	92.6	\pm	7.4	$&$	{--}	 	 	$\\
PG1151+117		&$	35.9	\pm	3.9	$&$	{--}	 	 	$&$	51.1	\pm	3.5	$&$	6.6	\pm	0.5	$\\
PG1202+281	&$	8.3	\pm	0.8	$&$	{--}	 	 	$&$	72.9	\pm	5.1	$&$	{--}	 	 	$\\
PG1211+143		&$	80.6	\pm	3.3	$&$	27.5	\pm	1.0	$&$	171.0	\pm	12.1	$&$	{--}	 	 	$\\
PG1216+069	&$	50.8	\pm	8.7	$&$	19.3	\pm	2.1	$&$	105.6	\pm	9.6	$&$	10.6	\pm	0.9	$\\
PG1226+023	&$	168.0	\pm	17.1	$&$	59.8	\pm	3.0	$&$	305.5	\pm	13.5	$&$	32.3	\pm	1.4	$\\
PG1229+204	&$	15.0	\pm	0.6	$&$	47.5	\pm	2.0	$&$	156.4	\pm	4.8	$&$	11.4	\pm	0.3	$\\
PG1244+026	&$	3.7	\pm	0.3	$&$	{--}	 	 	$&$	11.4	\pm	1.2	$&$	{--}	 	 	$\\
PG1259+593	&$	25.5	\pm	6.0	$&$	13.9	\pm	2.0	$&$	28.1	\pm	4.3	$&$	2.4	\pm	0.3	$\\
PG1302--102	&$	46.8	\pm	1.0	$&$	{--}	 	 	$&$	52.4	\pm	0.6	$&$	{--}	 	 	$\\
PG1307+085	&$	64.3	\pm	4.2	$&$	{--}	 	 	$&$	114.6	\pm	4.6	$&$	{--}	 	 	$\\
PG1310--108		&$	36.4	\pm	3.4	$&$	{--}	 	 	$&$	140.3	\pm	15.2	$&$	26.1	\pm	2.4	$\\
PG1322+659	&$	31.1	\pm	1.2	$&$	{--}	 	 	$&$	51.4	\pm	2.1	$&$	6.6	\pm	0.3	$\\
PG1341+258	&$	22.4	\pm	2.6	$&$	{--}	 	 	$&$	34.7	\pm	5.3	$&$	{--}	 	 	$\\
PG1351+695	&$	18.1	\pm	1.8	$&$	10.9	\pm	0.7	$&$	99.6	\pm	11.1	$&$	5.3	\pm	0.9	$\\
PG1352+183	&$	29.9	\pm	2.6	$&$	{--}	 	 	$&$	47.4	\pm	2.8	$&$	{--}	 	 	$\\
PG1402+261	&$	15.0	\pm	1.0	$&$	{--}	 	 	$&$	68.9	\pm	3.4	$&$	{--}	 	 	$\\
PG1404+226	&$	9.5	\pm	1.3	$&$	{--}	 	 	$&$	13.1	\pm	1.6	$&$	{--}	 	 	$\\
PG1415+451	&$	44.0	\pm	2.7	$&$	20.4	\pm	1.1	$&$	55.0	\pm	2.8	$&$	6.6	\pm	0.3	$\\
PG1416--129	&$	6.7	\pm	0.9	$&$	{--}	 	 	$&$	66.4	\pm	4.3	$&$	{--}	 	 	$\\
PG1425+267	&$	8.6	\pm	0.4	$&$	{--}	 	 	$&$	49.4	\pm	1.5	$&$	{--}	 	 	$\\
PG1426+015	&$	94.3	\pm	6.8	$&$	{--}	 	 	$&$	260.7	\pm	36.8	$&$	{--}	 	 	$\\
PG1427+480	&$	13.8	\pm	0.7	$&$	9.3	\pm	0.4	$&$	43.1	\pm	2.0	$&$	6.1	\pm	0.3	$\\
PG1434+590	&$	186.9	\pm	4.2	$&$	86.7	\pm	1.4	$&$	379.3	\pm	16.2	$&$	{--}	 	 	$\\
PG1435--067		&$	32.3	\pm	2.6	$&$	{--}	 	 	$&$	62.4	\pm	4.1	$&$	{--}	 	 	$\\
PG1440+356	&$	96.1	\pm	5.1	$&$	54.7	\pm	2.2	$&$	119.1	\pm	4.9	$&$	20.9	\pm	0.7	$\\
PG1444+407	&$	36.0	\pm	2.1	$&$	{--}	 	 	$&$	32.6	\pm	2.8	$&$	{--}	 	 	$\\
PG1448+273	&$	10.0	\pm	0.8	$&$	3.7	\pm	0.2	$&$	13.2	\pm	1.5	$&$	4.1	\pm	0.5	$\\
PG1501+106	&$	18.0	\pm	0.5	$&$	{--}	 	 	$&$	219.0	\pm	5.6	$&$	9.8	\pm	0.2	$\\
PG1512+370	&$	32.9	\pm	5.8	$&$	8.3	\pm	1.0	$&$	70.2	\pm	7.3	$&$	4.5	\pm	0.8	$\\
PG1519+226	&$	22.4	\pm	3.4	$&$	29.0	\pm	6.0	$&$	49.0	\pm	5.9	$&$	{--}	 	 	$\\
PG1534+580	&$	40.8	\pm	1.5	$&$	25.1	\pm	0.5	$&$	187.5	\pm	6.0	$&$	14.3	\pm	0.6	$\\
PG1543+489	&$	25.7	\pm	0.9	$&$	11.8	\pm	0.3	$&$	23.7	\pm	0.5	$&$	{--}	 	 	$\\
PG1545+210	&$	36.9	\pm	2.0	$&$	{--}	 	 	$&$	101.6	\pm	5.4	$&$	6.9	\pm	0.4	$\\
PG1552+085	&$	12.7	\pm	2.2	$&$	{--}	 	 	$&$	30.7	\pm	4.4	$&$	{--}	 	 	$\\
PG1612+261	&$	11.0	\pm	1.3	$&$	{--}	 	 	$&$	67.8	\pm	5.1	$&$	2.5	\pm	0.1	$\\
PG1613+658	&$	63.7	\pm	1.2	$&$	19.2	\pm	0.4	$&$	175.7	\pm	4.5	$&$	{--}	 	 	$\\
PG1617+175	&$	12.5	\pm	1.2	$&$	{--}	 	 	$&$	41.2	\pm	2.1	$&$	{--}	 	 	$\\
PG1626+554	&$	55.0	\pm	3.0	$&$	18.5	\pm	1.0	$&$	78.9	\pm	3.8	$&$	5.7	\pm	0.3	$\\
PG2112+059	&$	23.8	\pm	3.8	$&$	9.1	\pm	1.1	$&$	16.0	\pm	2.0	$&$	{--}	 	 	$\\
PG2130+099	&$	34.9	\pm	0.9	$&$	{--}	 	 	$&$	120.8	\pm	6.7	$&$	14.9	\pm	0.9	$\\
PG2214+139	&$	9.8	\pm	2.7	$&$	{--}	 	 	$&$	88.8	\pm	14.3	$&$	8.0	\pm	1.5	$\\
PG2233+134	&$	10.4	\pm	0.8	$&$	4.4	\pm	0.2	$&$	8.0	\pm	0.3	$&$	{--}	 	 	$\\
PG2251+113	&$	21.0	\pm	2.2	$&$	5.3	\pm	0.2	$&$	29.4	\pm	1.2	$&$	{--}	 	 	$\\
PG2304+042	&$	12.5	\pm	1.6	$&$	{--}	 	 	$&$	89.2	\pm	8.8	$&$	{--}	 	 	$\\
PG2308+098	&$	23.5	\pm	4.5	$&$	6.4	\pm	0.8	$&$	37.0	\pm	4.8	$&$	{--}	 	 	$
\enddata
\label{table:prop}
\tablecomments{
     Col. (1): Target ID. 
     Col. (2): line flux and error of N~{\sc v}. 
     Col. (3): line flux and error of Si~{\sc iv}+O~{\sc iv}. 
     Col. (4): line flux and error of C~{\sc iv}. 
     Col. (5): line flux and error of He~{\sc ii}.
}
\end{deluxetable}

%section 4
\section{result} \label{section:result}

%section 4.1
\subsection{Comparison of emission-line fluxes} \label{section:comparison}

In this study, we use three line flux ratios as metallicity indicators
among various diagnostics previously proposed, due to the limited data quality; 
namely, N~{\sc v}$\lambda$1240/C~{\sc iv}$\lambda$1549,
(Si~{\sc iv}$\lambda$1397+O~{\sc iv}]$\lambda$1402)/C~{\sc iv}$\lambda$1549,
and N~{\sc v}$\lambda$1240/He~{\sc ii}$\lambda$1640 (hereafter 
N~{\sc v}/C~{\sc iv}, (Si~{\sc iv}+O~{\sc iv}])/C~{\sc iv}, and N~{\sc v}/He~{\sc ii}, respectively).
Before using them as metallicity indicators in comparison with AGN properties,
here we first examine the correlation among the emission line fluxes.

Figure 5 (top panel) presents the comparison between the C~{\sc iv}$\lambda$1549 
and He~{\sc ii}$\lambda$1640 fluxes, which are used as a denominator of metallicity diagnostics. 
As photoionization models predict that the flux ratio of
these two lines does not depend on $Z_{\rm BLR}$ \citep[see, e.g., Figure 29 in]
[]{Nagao2006b}, a clear linear relation with a small scatter (0.20 dex) is 
present, showing that the flux ratio of these two lines is nearly constant. 
The constant flux ratio between C~{\sc iv} and He~{\sc ii} suggests that our flux measurements of the weak He~{\sc ii} line are reasonable although the He~{\sc ii} is difficult to fit due to blending with
Al~{\sc ii}.

In Figure 5 (bottom panel) we compare the N~{\sc v}$\lambda$1240 flux and 
the sum of the Si~{\sc iv}$\lambda$1397 and O~{\sc iv}]$\lambda$1402 fluxes,
which are used as a numerator of metallicity indicators. Again a clear 
positive correlation is present between them, with a somewhat larger 
scatter than that shown in the comparison between the C~{\sc iv} and He~{\sc ii} fluxes.
This larger scatter is partly caused by the fact that the N~{\sc v} flux depends 
on $Z_{\rm BLR}$ as well as the relative abundance of N~{\sc v} \citep[see, e.g.,][]{matsuoka2011b,Araki2012},
while the sum of the Si~{\sc iv} and O~{\sc iv}] fluxes mainly depends on $Z_{\rm BLR}$.

\begin{figure}
\includegraphics[width = 0.4\textwidth]{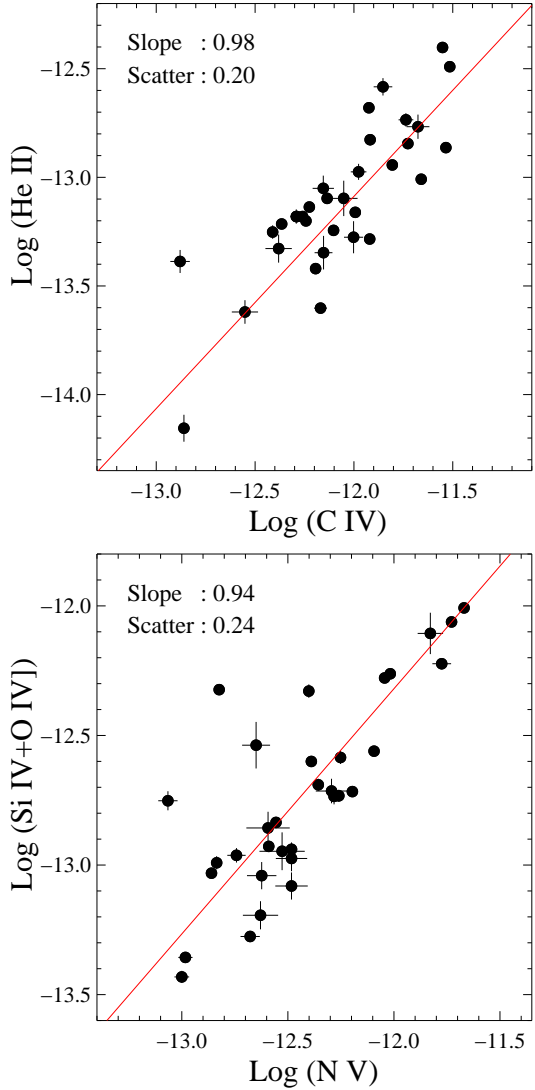}
\centering
\caption{
    Comparison of emission-line fluxes. {\it Top:} Comparison 
    between the He~{\sc ii}$\lambda$1640 and C~{\sc iv}$\lambda$1549 fluxes. 
    {\it Bottom:} Comparison between the sum of the Si~{\sc iv}$\lambda$1397 and
    O~{\sc iv}]$\lambda$1402 fluxes and N~{\sc v}$\lambda$1240 flux. 
    The derived Spearman's rank order correlation coefficients and their statistical significance	
    are $+0.795$, $3.2\times 10^{-8}$ (top) and $+0.751$,  $4.5\times 10^{-7}$ (bottom).
    \label{fig:allspec1}}
\end{figure}

%section 4.2
\subsection{Comparison among metallicity indicators} 
\label{section:comparison among metallicity indicators}

We compare 3 line flux ratios, namely,
N~{\sc v}/C~{\sc iv}, (Si~{\sc iv}+O~{\sc iv}])/C~{\sc iv}, and N~{\sc v}/He~{\sc ii},
as metallicity indicators adopted in this study.
As shown in Figure 6, the flux ratios of N~{\sc v}/C~{\sc iv} and N~{\sc v}/He~{\sc ii} 
present a clear correlation with a relatively small scatter, reflecting the constant
flux ratio between C~{\sc iv} and He~{\sc ii} (see Figure 5). 
The comparison between (Si~{\sc iv}+O~{\sc iv}])/C~{\sc iv} and N~{\sc v}/C~{\sc iv} 
also shows a relatively good correlation although a few outliers dominate
the scatter.
In the case of (Si~{\sc iv}+O~{\sc iv}])/C~{\sc iv} and N~{\sc v}/He~{\sc ii},
the comparison shows a less clear correlation, probably due to the larger
combined uncertainties on the flux measurements of weak lines (Si~{\sc iv}+O~{\sc iv}]\ and He~{\sc ii}).

\begin{figure}
\includegraphics[width = 0.4\textwidth]{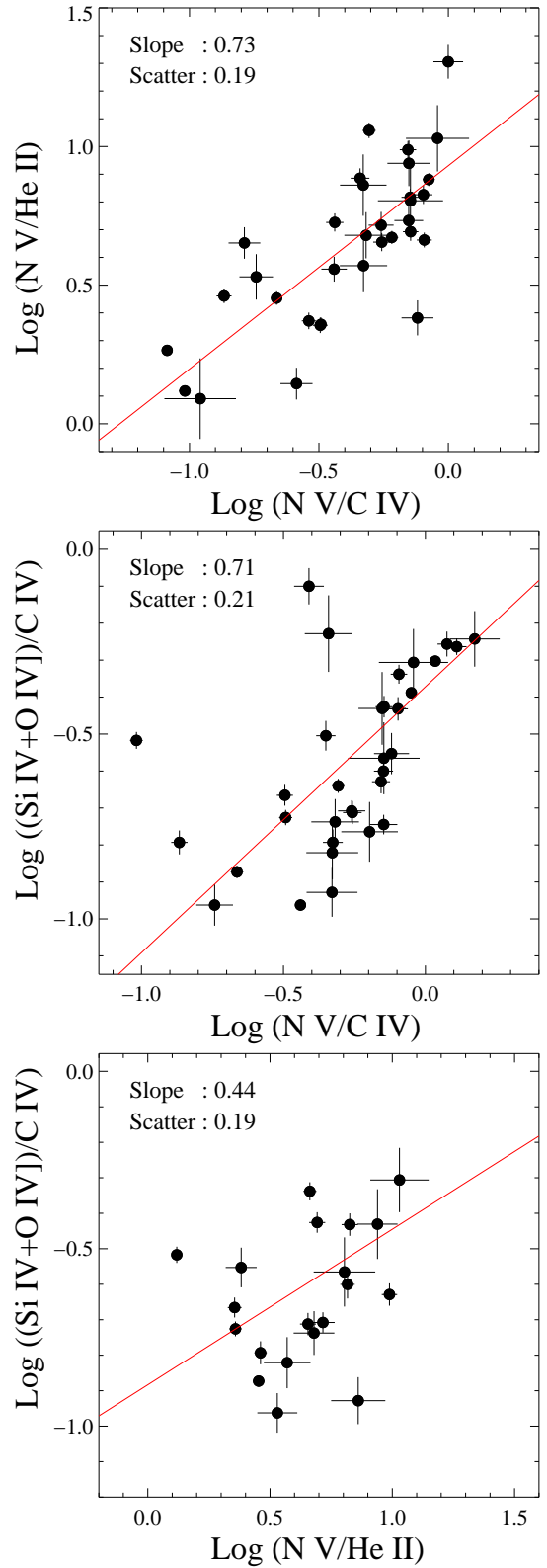}
\centering
\caption{
     The relation among three metallicity indicators. 
     Red line represents error weighted linear fit to the data. 
     The slopes of the linear fit and the data dispersion are
     shown at the upper-left corner in each panel.
     The derived Spearman's rank order correlation coefficients and their statistical significance	
    are $+0.726$, $2.0\times 10^{-6}$ (top), $+0.601$,  $2.1\times 10^{-4}$ (middle) 
    and $+0.329$,  $1.6\times 10^{-1}$ (bottom).
\label{fig:allspec1}}
\end{figure}

%section 4.3
\subsection{Comparison between metallicity and AGN properties} 
\label{section:comparison between metallicity and AGN properties}

In this section, we investigate the correlation between BLR metallicity inferred from
the emission line ratios and AGN properties, i.e., black hole mass, luminosity, 
and Eddington ratio, using the selected low-$z$ QSOs.
Figure 7 compares 3 metallicity indicators (N~{\sc v}/C~{\sc iv}, 
(Si~{\sc iv}+O~{\sc iv}])/C~{\sc iv}, and N~{\sc v}/He~{\sc ii}) with AGN 
properties ($M_{\rm BH}$, $L_{\rm bol}$, and $L_{\rm bol}/L_{\rm Edd}$), respectively.
A positive correlation is present between N~{\sc v}/C~{\sc iv} or 
N~{\sc v}/He~{\sc ii} with AGN luminosity, indicating a luminosity dependence of the BLR
metallicity, while it is less certain in the case of (Si~{\sc iv}+O~{\sc iv}])/C~{\sc iv},
probably due to the larger measurement uncertainty in Si~{\sc iv}+O~{\sc iv}] flux.
The apparent luminosity - BLR metallicity correlation of low redshift QSOs is similar to 
the results from previous studies based on high redshift AGNs \citep{Shemmer2004,Warner2004,
Nagao2006b}. 

We investigate which parameter between $M_{\rm BH}$ and $L_{\rm bol}$/$L_{\rm Edd}$
is more fundamental in driving the observed $L_{\rm AGN} - Z_{\rm BLR}$ relation. 
As for \mbh, there is no obvious \mbh\ dependence
on N~{\sc v}/C~{\sc iv} and (Si~{\sc iv}+O~{\sc iv}])/C~{\sc iv} while there is a 
possible positive correlation between N~{\sc v}/He~{\sc ii} and $M_{\rm BH}$.
These behaviors are in contrast to previous results obtained at high redshifts,
where significant positive correlations were reported between the metallicity
indicators and \mbh\ \citep{Warner2004,matsuoka2011b}.

On the contrary, clear positive dependences on $L_{\rm bol}$/$L_{\rm Edd}$
are present in N~{\sc v}/C~{\sc iv} and N~{\sc v}/He~{\sc ii}, while it is less
certain in the case of (Si~{\sc iv}+O~{\sc iv}])/C~{\sc iv}, due to the lack of 
objects in the range of $L_{\rm bol}$/$L_{\rm Edd}$ $<$ --1.5. For these low Eddington
ratio objects,  Si~{\sc iv}+O~{\sc iv}] lines are very weak and no secure measurements
are available.

In order to examine the statistical significance of these possible correlations,
we applied the Spearman's rank-order test to the data. The derived Spearman's 
rank-order correlation coefficients ($r_{\rm S}$) and their statistical significance 
($p$), which is the probability of the data being consistent with the null 
hypothesis that the flux ratio is not correlated with an AGN parameter, are 
given in Table 3.  The rank-order tests suggest that there are statistically 
significant positive correlations of N~{\sc v}/C~{\sc iv} with $L_{\rm bol}$ 
and $L_{\rm bol}$/$L_{\rm Edd}$, and a positive correlation of N~{\sc v}/He~{\sc ii} 
with $L_{\rm bol}$.
In contrast, all three diagnostic flux ratios show no statistically significant 
correlation with $M_{\rm BH}$.

\begin{figure*}
\includegraphics[width = 0.9 \textwidth]{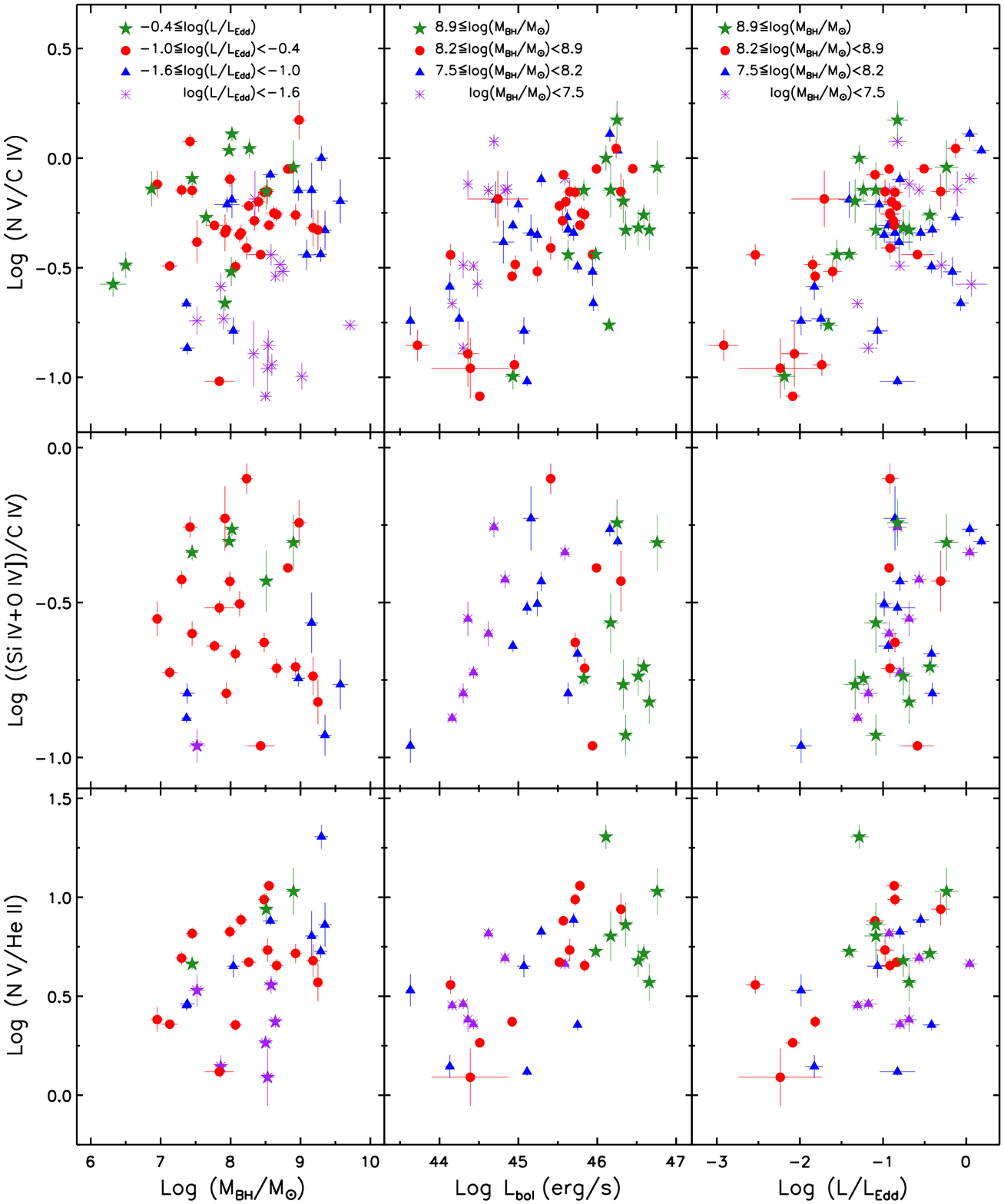}
\caption{
     The relation between metallicity indicators and AGN properties ($M_{\rm BH}$, 
     $L_{\rm bol}$, and $L_{\rm bol}/L_{\rm Edd}$). %The symbols represent the 
     %sample type; i.e., reverberation mass sample (circle) and single-epoch mass sample (triangle). 
     %The filled symbols denote higher quality data while
     %open symbols denote lower quality data (see Table 1). 
     The symbols and colors represent the 
     Eddington ratio or mass bin as indicated in the top panels.
\label{fig:allspec1}}
\end{figure*}

\begin{deluxetable}{lccc} 
\tablewidth{0pt}
\tablecolumns{4}
\tabletypesize{\scriptsize}
\tablecaption{Results of Spearman's rank-order correlation test}
\tablehead{
\colhead{Flux ratio} &
\colhead{$M_{\rm BH}$} &
\colhead{$L_{\rm bol}$} &
\colhead{$L_{\rm bol}/L_{\rm Edd}$}
\\
\colhead{(1)} & \colhead{(2)} & \colhead{(3)} & \colhead{(4)} 
}
\startdata
N~{\sc v}/C~{\sc iv} & 
     $r_{\rm S} = +0.076$ & 
     $r_{\rm S} = +0.487$ & 
     $r_{\rm S} = +0.473$ \\
  & $p = 5.3 \times 10^{-1}$ &
     $p = 2.1 \times 10^{-5}$ &
     $p = 3.6 \times 10^{-5}$ \\
(Si~{\sc iv}+O~{\sc iv}])/C~{\sc iv}  &
     $r_{\rm S} = -0.128$ & 
     $r_{\rm S} = +0.038$ & 
     $r_{\rm S} = +0.381$ \\
  & $p = 4.8 \times 10^{-1}$ &
     $p = 8.4 \times 10^{-1}$ &
     $p = 2.9 \times 10^{-2}$ \\
N~{\sc v}/He~{\sc ii}     &
     $r_{\rm S} = +0.417$ & 
     $r_{\rm S} = +0.622$ & 
     $r_{\rm S} = +0.311$ \\
  & $p = 1.6 \times 10^{-2}$ &
     $p = 1.1 \times 10^{-4}$ &
     $p = 7.8 \times 10^{-2}$ 
\enddata
\label{table:prop}
%\tablecomments{
%     Col. (1): Metallicity indicator. 
%     Col. (2): Spearman correlation coefficent and significance between 
%     metallicity indicators and black hole mass 
%     Col. (3): Spearman correlation coefficent and significance between 
%     metallicity indicators and AGN bolometric luminosity 
%     Col. (4): Spearman correlation coefficent and significance between 
%     metallicity indicators and Eddington ratio.
%}
\end{deluxetable}

%section 5
\section{Discussion}\label{section:Discussion}
\begin{figure}
\includegraphics[width = 0.47\textwidth]{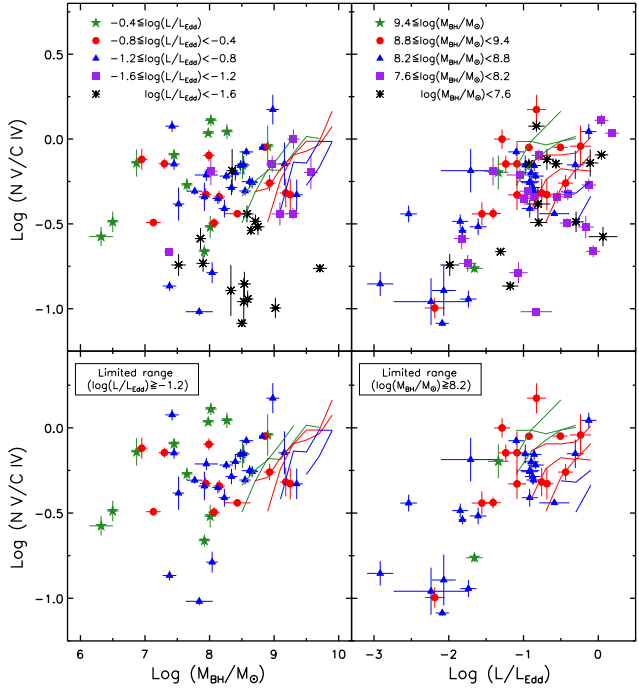}
\caption{{\it Top:}
     Metallicity indicator (N~{\sc v}/C~{\sc iv} flux ratio) 
     as functions of $M_{\rm BH}$ (left) and $L/L_{\rm Edd}$ (right).
     Low-$z$ QSOs are represented with various symbols 
     while high-$z$ QSOs are denoted 
     by solid lines (\citealt{matsuoka2011b}). 
     The colors represent different mass and Eddington ratios
     as indicated in the upper panels.
     {\it Bottom:} Metallicity indicator vs. 
     $M_{\rm BH}$ (left) and $L/L_{\rm Edd}$ (right), after excluding low \mbh\
     and low Eddington AGNs from low-$z$ sample for proper comparison.
\label{fig:allspec1}}
\end{figure}

We compare our results obtained for AGNs at $z < 0.5$ with the previous results 
obtained for high-$z$ QSOs (z$\sim$2.5; \citealt{matsuoka2011b}), in order to investigate 
possible redshift evolution and the origin of the metallicity scaling relations in AGNs.
In Figure 8, we overplot the metallicity indicators as functions of $M_{\rm BH}$ and
$L/L_{\rm Edd}$ for high-$z$ QSOs adopted from \citep{matsuoka2011b},  
along with the measurements of low-$z$ QSOs.
The low-$z$ objects are more dispersed than high-$z$ objects, partly 
because of the larger measurement uncertainties of the low-$z$ objects. 
Note that the emission line fluxes were measured for individual objects in the low-$z$ sample,
while the composite spectra of high-$z$ QSOs in each black hole mass and Eddington ratio bin were 
used for the emission line flux measurements \citep[see][]{matsuoka2011b}.

Interestingly, the low-$z$ and high-$z$ samples appear to show different trends in Figure 8 (top panels). 
As mentioned in \S 4.3, the low-$z$ sample shows only 
marginal correlations between metallicity indicators and $M_{\rm BH}$, while
the high-$z$ sample clearly shows positive correlations of metallicity indicators with \mbh.  
On the other hand, the metallicity indicators of the low-$z$ sample show much 
stronger positive correlations with $L/L_{\rm Edd}$ than with $M_{\rm BH}$, 
while the correlations between metallicity indicators and $L/L_{\rm Edd}$ are 
less evident in the high-$z$ sample.

The difference in the trend with Eddington ratios between low-$z$ and high-$z$ samples
is partly caused by the much wider range of $L/L_{\rm Edd}$ 
($-3 < {\rm log} [L/L_{\rm Edd}] < 0$) covered by the low-$z$ sample than 
that covered by the high-$z$ sample ($-1 < {\rm log} [L/L_{\rm Edd}] < 0$). 
As shown in Figure 8, low-$z$ AGNs with 
${\rm log} [L/L_{\rm Edd}] < -1.5$ (that is not covered in the high-$z$ sample) show systematically 
lower N~{\sc v}/C~{\sc iv} flux ratios, leading to the more evident dependence
on $L/L_{\rm Edd}$. Therefore, our results do not necessarily suggest that
the $L/L_{\rm Edd}$ dependence on the metallicity indicators is systematically
different between low-$z$ and high-$z$ QSOs. For proper comparison, low Eddington AGNs
(${\rm log} [L/L_{\rm Edd}] < -1.5$) are required at high-$z$.
These results imply that the accretion activity of black holes are closely 
related with metal enrichment at the central part of host galaxies.

The dependence of metallicity indicators on $M_{\rm BH}$ appears to be
different between the low-$z$ and high-$z$ samples. 
At high-$z$, more massive AGNs have higher metallicity although the $M_{\rm BH}$ range is
small (8.5 $<$ log \mbh $<$ 10). In contrast, the low-$z$ sample shows much larger scatter
without clear trend between \mbh\ and metallicity indicators.
However, the large scatter in the low-$z$ sample is partly 
caused by the low Eddington AGNs, which have systematically lower N~{\sc v}/C~{\sc iv}
flux ratios, as described as the metallicity-Eddington ratio relation. 
However, if AGNs with similar Eddington ratios are selected (e.g., same color objects in Figure 8), 
then there appears to be a weak trend of metallicity with \mbh, suggesting that 
at low-$z$, \zBLR\ has weak dependency on \mbh\ at fixed Eddington ratios. 

For high-$z$ QSOs, the tightness of the observed $Z_{\rm BLR}-M_{\rm BH}$ correlation is 
probably caused by the limited range of the Eddington ratios since only high Eddington
QSOs were included in the high-$z$ sample. However, inclusion of low Eddington AGNs
will presumably weaken the correlation as in the case of the low-$z$ sample. 
As a consistency check, we match the ranges of \mbh\ and Eddington ratio between 
high-$z$ and low-$z$ samples, 
by excluding low \mbh\ and low Eddington objects from the low-$z$ sample as presented
in the bottom panels of Figure 8.
As expected, both high-$z$ and low-$z$ samples in the matched dynamical range show similar 
metallicity dependency on both \mbh\ and Eddington ratio although the \mbh-\zBLR\ relation 
is much weaker in low-$z$ than at in high-$z$.

Our results imply that the BLR metallicity of low-$z$ AGNs mainly depends on the Eddington ratio,
and weakly depends on \mbh. Currently, it is unknown how \zBLR\ scales with average
gas metallicity of host galaxies. Nevertheless, the observed metallicity dependency 
on \mbh\ may imply that there is connection among BH growth, gas enrichment, and
galaxy evolution. 

Assuming that \zBLR\ correlates with gas metallicity of host galaxies, we try to
understand the observed relations with several scenarios.   
The correlation between \mbh\ and \zBLR\ at high-$z$ can be interpreted as a consequence of 
the combination of the galaxy mass-metallicity 
relation and the $M_{\rm BH}-M_{\rm bulge}$ relation \citep{Warner2003,matsuoka2011b}.
For QSOs at $z \sim 2-3$ that corresponds to the peak of the quasar activity 
in the cosmological timescale, it has been often claimed that the major merger triggers
the AGN activity through the efficient mass fueling onto black holes \citep{Hasinger2008,Li2010}. Here
the major merger event reduces the angular momentum of gas clouds at the nuclear
regions of the quasar host galaxies, providing efficient mass fueling. In this case,
the metallicity of accreting gas onto the nucleus may be characterized by the mass-metallicity
relation of the host galaxy. Regarding the black hole to host galaxy connection,
the $M_{\rm BH}-M_{\rm bulge}$ relation has not been observationally defined
at high-$z$ although several studies indicated that \mbh-to-$M_{\rm bulge}$ ratio increases
with redshift \citep[e.q.,][]{Woo2006,Woo2008,Bennert2010,Bennert2011,Merloni2010}.
Thus, if we assume a scaling relation between \mbh\ and galaxy mass, presumably 
with higher normalization than the local \mbh-$M_{\rm bulge}$ relation,
then black hole mass scales with both stellar mass and gas metallicity, hence 
the \mbh-\zBLR\ relation is naturally expected at high-$z$. 

The weaker correlation of \mbh\ with \zBLR\ at low-$z$ can be interpreted as
combination of two additional effects. First, gas metallicity increased from high-$z$ to low-$z$
by further star formation after major BH growth. In other words, for the same galaxy mass 
(or black hole mass), metallicity has been increased, particularly at the nuclear region, 
leading to higher metallicity at fixed \mbh\ compared to high-$z$ objects.
For example, nuclear star formation induced by secular process (e.g., bar instability)
%\citep{Davies2007,Wild2010} 
and galaxy interaction may sufficiently increase metallicity \citep[e.q.,][]{Maiolino2008}. 
Another difference between low-$z$ and high-$z$ is the gas fraction, especially for 
massive galaxies like quasar host galaxies. The minor merger process is more 
important as an AGN triggering mechanism at low-$z$ \citep[e.q.,][]{Taniguchi1999,Cisternas2011}.
Thus, as a consequence of minor mergers between a gas-poor massive galaxy (i.e., 
a quasar host galaxy) and a relatively gas-rich less-massive galaxy, the metallicity of 
accreting gas onto the nucleus can be largely affected by the chemical property of 
the merging, less-massive galaxy. In this case, 
the metallicity of accreting gas does not simply reflect the mass-metallicity relation
of the host galaxy. 
Second, the \mbh-to-$M_{\rm bulge}$ ratio may change over cosmic time. 
If black hole mass was higher
at fixed stellar mass at high-$z$, then by combining galaxy mass-metallicity relation
and $M_{\rm BH}-M_{\rm bulge}$ relation, we may expect similar \mbh-metallicity relation 
in low-$z$, but with a different normalization. 
In other words, at fixed metallicity, \mbh\ is lower in low-$z$ than in high-$z$.    
Although it is beyond the scope of the current work to quantify and compare these two effects,
it is reasonable to conclude that coupling between black hole mass with gas metallicity becomes much weaker with decreasing redshift.

%However at low-$z$, the minor merger process is more 
%important as a triggering mechanism \citep[e.q.,][]{Taniguchi1999,Cisternas2011}.
%Another difference between low-$z$ and high-$z$ is the gas fraction, especially for 
%massive galaxies like quasar host galaxies. As a consequence of minor mergers between
%a gas-poor massive galaxy (i.e., a quasar host galaxy) and a relatively gas-rich 
%less-massive galaxy, the metallicity of accreting gas onto the nucleus is largely affected
%by the chemical property of the merging, less-massive galaxy. In this case, 
%the metallicity of accreting gas is not determined by the $M_{\rm galaxy}-M_{\rm BH}$ 
%relation of the host galaxy. This is actually reported observationally at low redshifts; i.e., 
%massive galaxies with signs of merger events show lower gas metallicities than the 
%expectations from the $M_{\rm galaxy}-M_{\rm BH}$ relation \citep{SolAlonso2010}.
%Therefore we do not see the $Z_{\rm BLR}-M_{\rm BH}$ correlation in those low-$z$ 
%QSOs, that are triggered by minor mergers.

\section{Summary \& Conclusion}\label{section:Conclusion}
To investigate the chemical properties of low-$z$ QSOs, we measured the flux ratios of 
the rest-frame UV emission lines as metallicity indicators using a sample of 70 low-$z$ PG QSOs 
at $z < 0.5$. By comparing BLR gas metallicity with black hole mass, luminosity and 
Eddington ratio,
we find that \zBLR\ correlates with Eddington ratio while \zBLR\ shows much weaker
correlation with \mbh, indicating that the metal enrichment at the central part of
host galaxies is closely connected to the accretion activity of AGN. 
These trends are different from high-$z$ QSOs, which shows 
a tighter correlation between \zBLR\ and \mbh\ and a weaker correlation between \zBLR\ and 
Eddington ratio. The apparent difference between low-$z$ and high-$z$ samples 
seems to be caused by the limited dynamical range in the high-$z$ sample.
Various star formation mechanism can increase BLR gas metallicity the cosmic
time, increasing the scatter in the metallicity correlation with properties
of AGN in low-$z$.

\acknowledgements
This work was supported by the Korea Astronomy and Space Science Institute (KASI) grant funded by the Korea government (MEST).
J.H.W acknowledges the support by the National Research Foundation of Korea (NRF) grant funded by the Korea government (MEST) (No. 2012-006087).
T.N. acknowledges the support by JSPS (grant no. 23654068)
and by the Hakubi project in Kyoto University.
The Mikulski Archive for Space Telescopes (MAST) is a NASA funded project to support and provide to the astronomical community 
a variety of astronomical data archives, with the primary focus on scientifically related data sets in the optical, ultraviolet, and near-infrared 
parts of the spectrum. MAST is located at the Space Telescope Science Institute (STScI).

%\bibliography{metallicity}

\end{document}